\documentclass[prl, twocolumn,preprintnumbers,aps]{revtex4}
\usepackage{graphicx}
\usepackage{makeidx}

\usepackage{color}

\newcommand{\be}{\begin{eqnarray}}
\newcommand{\ee}{\end{eqnarray}}

\begin{document}

\title{Dissipation of Modified Entropic Gravitational Energy Through Gravitational Waves}

\author{Clovis Jacinto de Matos$^{\rm 1}$}

\affiliation{$^{\rm 1}$European Space Agency, 8-10 rue Mario
Nikis, 75015 Paris, France}

\date{22 September 2011}

\preprint{}

\begin{abstract}
The phenomenological nature of a new gravitational type
interaction between two different bodies derived from Verlinde's
entropic approach to gravitation in combination with Sorkin's
definition of Universe's quantum information content, is
investigated. Assuming that the energy stored in this entropic
gravitational field is dissipated under the form of gravitational
waves and that the Heisenberg principle holds for this system, one
calculates a possible value for an absolute minimum time scale in
nature $\tau=\frac{15}{16} \frac{\Lambda^{1/2}\hbar
G}{c^4}\sim9.27\times10^{-105}$ seconds, which is much smaller
than the Planck time $t_{P}=(\hbar G/c^5)^{1/2}\sim
5.38\times10^{-44}$ seconds. This appears together with an
absolute possible maximum value for Newtonian gravitational forces
generated by matter $F_g=\frac{32}{30}\frac{c^7}{\Lambda \hbar
G^2}\sim 3.84\times 10^{165}$ Newtons, which is much higher than
the gravitational field between two Planck masses separated by the
Planck length $F_{gP}=c^4/G\sim1.21\times10^{44}$ Newtons.
\end{abstract}


\maketitle

\section{Modified Entropic Gravitation Between Two Bodies}

In a recent paper \citep{dematos}, the author derived a new
gravitational type force law, Equ.(\ref{e0}), starting from
Verlinde's entropic approach to gravitation, and assuming Sorkin's
hypothesis that the total amount of information in the Universe is
directly proportional to the Universe four-Volume.
\begin{equation}
F=3 \frac{\Lambda ^{1/2} G^2 \hbar}{c^3} \frac {M_0 m_0}{R^3}\sim
5.93\times 10^{-106} \frac {M_0 m_0}{R^3}\label{e0}
\end{equation}
Where $G$ is Newton's universal gravitational constant, $\hbar$ is
the Planck constant divided by $2\pi$, $\Lambda=1.29\times
10^{-52} [m^{-2}]$ is the cosmological constant \citep{Spergel},
and $M_0$, $m_0$ are the respective masses of the interacting
bodies whose center of mass are separated by the distance $R$.
Since this force contains a proportionality constant $\Lambda
^{1/2} G^2 \hbar /c^3\sim 5.93\times 10^{-106}$ which is extremely
small (to say the least), It is easy to conceive that this new
type of gravitational force has never been experimentally detected
in the context of the gravitational interaction between massive
bodies. From the force law Equ.(\ref{e0}) one deduces that the
total mechanical energy stored in a gravitating binary system
orbiting under the single influence of this force is:
\begin{equation}
E=\pm \frac{3}{2} \frac{\Lambda ^{1/2} G^2 \hbar}{c^3} \frac {M_0
m_0}{R^2}\label{e1}
\end{equation}
For the moment one does not make any assumption with respect to
the attractive or repulsive character of the force $F$. This
explains why we consider two possible signs for the total
mechanical energy of the system (negative if the force is
attractive, positive if the force is repulsive).

\section{Gravitational Radiation Emitted by a Binary
System under the Single Influence of a Modified Entropic
Gravitational Force}

Let one consider that the two bodies have identical masses,
$M_0=m_0$, and that they are spinning with angular velocity
$\omega$ about the center of mass of the binary system. Thus they
form a kind of spinning dumbbell. General relativity predicts that
this system will emit gravitational waves, with radiating power
$\wp$ \citep{Forward}.
\begin{equation}
\wp=\frac{8}{5}\frac{G}{c^5}m_0^2 R^4 \omega^6 \label{e2}
\end{equation}
Where $R$ is the distance between the respective centers of mass
of the two bodies.

Dividing Equ.(\ref{e1}) by Equ.(\ref{e2}) one deduces the interval
of time $\tau$ required to dissipate entirely the total mechanical
energy of the system under the form of gravitational radiation:
\begin{equation}
\tau=\frac{E}{\wp}=\frac{15}{16} \Lambda^{1/2} c^2 \hbar G
(R\omega)^{-6}\label{e3}
\end{equation}
Since $\omega=v/R$ ($v$ being the tangential velocity of
rotation), and assuming the asymptotic limit $v=c$, one deduces
from Equ.(\ref{e3}) a dissipating time $\tau$ which does not
depend on the relative distance $R$ between the two bodies.
\begin{equation}
\tau=\frac{E}{\wp}=\frac{15}{16} \frac{\Lambda^{1/2}\hbar
G}{c^4}\sim9.27\times10^{-105} [Seconds].\label{e4}
\end{equation}
Note that the masses of the bodies $m_0$ have disappeared from
Equ.(\ref{e3}) and Equ.(\ref{e4}) due to the principle of
equivalence.

 \section{Dicussion and Conclusions}

Assuming that the the time required to dissipate the mechanical
energy of the binary system, Equ.({\ref{e1}),should also comply
with the Heisenberg uncertainty principle, one calculates an
alternative dissipation time $\tau'$
\begin{equation}
\tau'\sim\frac{\hbar}{E}=\frac{2}{3}\frac{c^3}{\Lambda^{1/2}G^2}\frac{R^2}{m_0^2}
\label{e5}
\end{equation}
Imposing that both decay times must be equal to each other,
$\tau=\tau'$, one deduces a maximum Newtonian binding
gravitational force for the binary system.
\begin{equation}
F_g=G\frac{m_0^2}{R^2}=\frac{32}{30} \frac{c^7}{\Lambda \hbar
G^2}\sim3.84\times 10^{165} [Newtons] \label{e6}
\end{equation}
Neither the time interval $\tau$, and the gravitational Newtonian
force $F_g$ depend on the radius $R$ of the system or on the
angular frequency $\omega$ at which the two bodies rotate around
their commune center of mass. Instead both $\tau$ and $F_g$ depend
only on the fundamental constants $G, \hbar, c, \Lambda$. Since
the value of $\tau$ is much smaller than the Planck time,
$t_{P}=(\hbar G/c^5)^{1/2}\sim 5.38\times10^{-44}$ seconds, it is
tempting to consider this interval of time as an absolute minimum
time in nature. A similar argument could also hold for the
gravitational Newtonian force $F_g$, which could be understood as
a maximum scale value (for this type of force) since it is much
higher than the Newtonian gravitational force between two Planck
masses separated by the planck length
$F_{gP}=c^4/G\sim1.21\times10^{44}$ Newtons.

It is also instructive to note that the Planck time $t_p$
corresponds to a good approximation (within $85\%$ agreement) to
the geometric mean between the Universe age $T_U$ and the decay
time $\tau$
\begin{equation}
t_p\sim\sqrt{\tau T_U}\label{e7}
\end{equation}
where $T_U\sim1/H_0=4.348\times10^{17} s$ ($H_0$ being the Hubble
constant). This encourages one interpreting the time $\tau$ as the
minimum possible time in nature and the Universe age $T_u$ as the
maximum possible time with physical meaning.

\section{Acknowledgements}

\acknowledgments{I would like to thank the referee who reviewed my
paper on "modified entropic gravitation in superconductors"
\cite{dematos}, for outlining that the numerical value of the
fundamental constant appearing in the force law between two
different universes: $3 \Lambda ^{1/2} G^2 \hbar /c^3\sim
5.93\times 10^{-106} \ [m^4.s^{-2}.Kg^{-1}]$ is extremely close to
the square of the proportionality constant appearing in the
relativistic law determining the power radiated by an accelerated
physical system: $(G/c^5)^2=7.58\times 10^{-106}$
$[s^3.m^{-2}.Kg^{-1}]$. The author would like also to thank the
referee of the present paper for precious guidance in the physical
interpretation of the physical phenomena presented in this paper.}

\end{document}